\documentclass{article}
\usepackage{graphicx}
\usepackage{epsfig}

\begin{document}

\begin{center}
{\bfseries Reconstruction of the optical potential from scattering data}

\vskip 5mm

N.A. Khokhlov$^{\dag}$, V.A. Knyr$^{\dag \dag}$

\vskip 5mm

{\small  {\it Khabarovsk State University of Technology, 680035
Khabarovsk, Russia }
\\
$\dag$ {\it E-mail: khokhlov@fizika.khstu.ru }; $\dag\dag$ {\it
E-mail: knyr@fizika.khstu.ru }}
\end{center}

\vskip 5mm

\begin{abstract}
We propose a method for reconstruction of the optical potential from scattering data. The algorithm is a two-step
procedure. In the first step the real  part of the potential is determined analytically via solution of the Marchenko
equation. At this point we use a diagonal Pad\'{e} approximant of the corresponding unitary $S$-matrix. In the second
step the imaginary part of the potential is determined via the phase equation of the variable phase approach. We assume
that the real and the imaginary parts of the optical potential are proportional. We use the phase equation to calculate
the proportionality coefficient. A numerical algorithm is developed for a single and for coupled partial waves. The
developed procedure is applied to analysis of $^{1}S_{0}$ $NN$, $^{3}SD_{1}$ $NN$,  $P31$ $\pi^{-} N$ and $S01$
$K^{+}N$ data.
\end{abstract}

\vskip 10mm

\section{\textbf{Introduction}}

A lot of developments and applications of the classical approaches of Gelfand, Levitan \cite{Gelfand} and Marchenko
\cite{Marchenko} for the solution of the inverse-scattering problem at fixed angular momentum exist, and we have
several of excellent reviews on the subject \cite{Newton, Chadan}. The direct application of these approaches to
construction of local two-body potentials which are phase equivalent to the effective potentials occurring in theories
describing reactions of composite particles is impossible. For, such potential must be complex in order to reproduce
the loss of flux above the inelastic threshold. But it must reproduce the real phase shifts below this threshold and
must be real itself. These requirements are incompatible for potentials being energy independent by construction. For
very low threshold these approaches are applicable and produce energy independent complex potentials
\cite{Papastylianos, Alt}. In the general case the empirical energy dependent optical potentials are usually inferred
by fitting of the parameters of an assumed analytic potential \cite{Green,Geramb}. This approach has two major
shortcomings: a complexity and inconvenience of fitting simultaneously many nonlinear parameters; and lack of
correlation of the parameters obtained at various energies.

In this paper we develop an inversion method that is free of these shortcomings. The method is based on a fixed-$l$
inverse scattering theory and on a special parameterization of the optical potential. The proposed procedure is a
two-step process. In the first step the phase shift data are used to determine a real potential via solution of the
Marchenko equation. At this point we use a diagonal Pad\'{e} approximant $[M/M]$ of the corresponding unitary
$S$-matrix. In the second step the imaginary part of the potential is determined via the phase equation of the variable
phase approach \cite{Calo}. We assume that the real and the imaginary parts of the optical potential are proportional.
The value of the proportionality coefficient is predicted by the phase equation and is refined by the iterative
algorithm. We develop this method for single and for coupled partial waves. The whole procedure is applied to analyze
$^{1}S_{0}$ $NN$, $^{3}SD_{1}$ $NN$ data and to analyze $P31$ $\pi-N$ and $S01$ $K+N$ data.  These analyses demonstrate
that prediction for the proportionality coefficient from the phase equation is very close to a precise value that
reproduce the experimental loss of flux.

The plan of the paper is as follows. In Sect. 2 we describe the inverse scattering techniques based on the Marchenko
integral equation. The used diagonal Pad\'{e} approximants of the corresponding $S$-matrix allow an analytical solution
of the Marchenko integral equation \cite{Marchenko, Chadan}. For single partial wave the general solution was presented
in \cite{Papastylianos, Kirst}. We present a solution for coupled partial waves. These techniques produce real local
potentials from phase shift analysis data. In Sect. 3 we consider the phase equation. We investigate how the $S$-matrix
is changing with certain change of the potential.   This consideration shows advantages of proportionality of the real
and imaginary parts of the optical potential. In Sect. 4 the feasibility of the method is shown in the examples of
analyses of $NN$, $\pi^{-}N$ and $K^{+}N$ scattering data.

\section{Inversion algorithm}

 The Marchenko inverse scattering theory is viewed in detail
 in Refs. \cite{Marchenko, Newton, Chadan}. We shall, therefore,
 only briefly describe this formalism.

 The input data of the Marchenko inversion are
\begin{equation}
\label{eq1} \left\{ {S\left( q \right), \left( { 0 < q < + \infty
} \right),\mbox{  }\tilde{q}_{j},\mbox{  }M_{j},\mbox{  }j =
1,...,n_{\mbox{\tiny{b}}} } \right\}, \label{initial_for_Mar}
\end{equation}

\noindent where $S\left( q \right)$ is the scattering matrix dependant on the relative momentum $q$, $q^2 = Em$,
$\tilde{q}_j ^2 = mE_j \le 0, \quad E_j $ is the energy of the j-th bound state, so that $\imath \tilde{q}_j \ge 0$,
$m$ is the reduced mass. The $M_j$ matrices give the asymptotic behavior of the corresponding normalized bound states.

We proceed from the Marchenko equation for single channel
\begin{equation}
\label{eq2} F(x,y) + L(x,y) + \int\limits_x^{ + \infty }
{L(x,t)F(t,y)dt} = 0, \label{Mar}
\end{equation}

\noindent where the input kernel is given by

\begin{equation}
\label{eq3} F\left( {x,y} \right) = \frac{1}{2\pi }\int\limits_{ -
\infty }^{ + \infty } {h_l^{+} \left( {qx} \right)\left( {I -
S\left( q \right)} \right)} h_l^{+}\left( {qy} \right)dq +
\sum\limits_{j = 1}^{n_{\mbox{\tiny{b}}} } {M_j^2 } h_l^{+} \left(
{iq_j x} \right)h_j^{+} \left( {iq_j y} \right),
\end{equation}

\noindent  $h_l^ + \left( z \right)$ are the Riccati-Hankel
functions.

The output kernel $L\left( {x,y} \right)$ gives the reconstructed
potential
\begin{equation}
\label{eq4} V\left( r \right) = - \frac{dL\left( {r,r}
\right)}{dr}.
\end{equation}

This local energy independent operator $V\left(r\right)$  links
the Marchenko equation (\ref{Mar}) and the radial Schr\"{o}dinger
equation of a fixed angular momentum,
\begin{equation}
\left[-\frac{d^2}{dr^2}+\frac{l(l+1)}{r^2}+V(r)\right]\psi(r,q)=q^2\psi(r,q).
\label{Schr}
\end{equation}
The scattering matrix $S\left( q \right)$, matrices $M_j$ and
energies $E_j$ are the output data of the direct scattering
problem associated with the Schr\"{o}dinger equation (\ref{Schr}).

It has been known for several decades now that $S$ matrices rational in $q$ (ratio of polynomials) correspond to
potentials known as Bargmann potentials expressible in terms of the elementary functions \cite{Marchenko, Newton,
Chadan}. Such fraction may have the same truncated Taylor series as the $S$ matrix it represents. It is then called a
Pad\'{e} approximant. Conjectures and theorems concerning the convergence and analytic continuation properties of
Pad\'{e} approximants  are collected in \cite{Pade}.
 For single partial wave the general solution of the Marchenko
equation via Pad\'{e} approximant  of the $S$-matrix was presented in \cite{Papastylianos} and in \cite{Kirst}. We
shall, therefore,
 only present it briefly and turn to the case of coupled partial waves.

A diagonal Pad\'{e} approximant of the $S$-matrix is given by

\begin{equation}
\label{eq5a} S\left( q \right) =e^{2\imath \delta}= \frac{f_2
\left( q \right) - \imath f_1 \left( q \right)}{f_2 \left( q
\right) + \imath f_1 \left( q \right)} \quad
\end{equation}

$f_1 \left( q \right)$ and $f_2 \left( q \right)$ are an odd and even polynomials of $q$, which do not turn to zero at
the real axis simultaneously.

This approximant leads to the following expression for the phase shifts $\delta \left( q \right)$

\begin{equation}
\label{eq6a} tg\left( { - \delta \left( q \right)} \right) =
\frac{f_1 \left( q \right)}{f_2 \left( q \right)}
\end{equation}

Inasmuch as $\lim_{q\rightarrow \infty}\delta(q)\sim 1/q$ for
regular potentials, it is evident that degree of $f_{1}(q)$ must
be less than degree of $f_{2}(q)$ by 1.
Let us select $N$ discrete momenta $q_{i}$ such that the corresponding $\delta \left( q_{i} \right)= \delta _i $ are
known. Use of these values in eq. (\ref{eq6a}) transforms the latter into a set of inhomogeneous linear equations from
which $N$ coefficients of the polynomials $f_1 \left( q \right)$ and $f_2 \left( q \right)$ can be determined. This is
a usual strategy \cite{Papastylianos}, but since any set of $q_{i}$ is experimentally limited from above
($q_{i}<q_{max}$) there is some uncertainty in determination of $S(q)$. Even if a set of $q_{i}$ is dense and the
agrement between the data and the used approximant is excellent
the arbitrary behavior of $S(q)$ above $q_{max}$ guarantees that a solution of the inverse problem is arbitrary as
well. We assume that about and above $q_{max}$ the true $\delta(q)$ depends only slightly on
 details of the potential and does not depend on its asymptotic at $r>r_{max}$. Then we take a model potential (in our
calculations $V_{model}(r)=A\exp(-br)$) and fit parameters ($A$ and $b$) so that
\begin{equation}
\delta_{model}(\approx q_{max})\approx\delta(\approx q_{max}),
\end{equation}
here signs $\approx$ mean that about $q=q_{max}$ the chosen model potential gives a phase curve which goes inside error
bars. This means that we take $\delta_{model}(q)$ as an asymptotic for $\delta(q)$ when $q>q_{max}$. In the segment
$[0,q_{max}]$ the $\delta(q)$ is defined by some spline that approximates the data points $\{q_{i},\,\delta_{i}\}$. In
this way we may control the Pad\'{e} fit in the line segment $[0,Q_{max}]$, where $Q_{max}$ is arbitrary and is fixed
by convergence of the whole inversion procedure. The needed accuracy of approximant is attained by increasing of $N$
which in turn defines degrees of $f_{1}(q)$ and $f_{2}(q)$.

Approximant (\ref{eq6a}) leads to a degenerate input kernel $F\left( {x,y} \right)$. We calculate the integral in eq.
(\ref{eq3}) using the residue theorem. For approximant (\ref{eq5a}) the result of the integration is
\begin{equation}
\begin{array}{c}
\label{eq7} F\left( x,y \right) =\left.\imath\sum\limits_{i =
1}^{n_{\mbox{\tiny{pos}}}} Res \left[{h_l^{+} \left( {qx}
\right)\left( {I - S\left( q \right)} \right)} h_l^{+}\left( {qy}
\right)\right] \right|_{q=\beta_i}
 + \sum\limits_{i =
1}^{n_{\mbox{\tiny{b}}} } {M_i^2 } h_l^{+} \left( {\tilde{q}_i x}
\right)h_l^ + \left( {\tilde{q}_i y} \right)  =\\
=\sum\limits_{i = 1}^{n_{\mbox{\tiny{pos}}}} {b_i h_l^ + \left(
{\beta _{i} x} \right)h_l^{+} \left( {\beta _{i} y}
\right)}+\sum\limits_{i = 1}^{n_{\mbox{\tiny{b}}} } {M_i^2 } h_l^
+ \left( {\tilde{q}_i x} \right)h_l^ + \left( {\tilde{q}_i y}
\right)=
\sum\limits_{j = 1}^n {b_j h_l^ + \left( {\beta _j x} \right)h_l^
+ \left( {\beta _j y} \right)} ,
\end{array}
\end{equation}

 \noindent where $\beta _i $ ($i =
1,...,n_{\mbox{\tiny{pos}}} )$ are all $S$-matrix poles with
$\Im\, \beta_i>0$,
$\beta=\{\beta_{1},..\beta_{n_{pos}},\tilde{q_{1}},...\tilde{q_{n_{b}}}\}$,
$n=n_{pos}+n_{b}$. We assume that all poles are of first order so
that
\begin{equation}
\begin{array}{c}
\left.Res \left[{h_l^{+} \left( {qx} \right)\left( {I - S\left( q
\right)} \right)} h_l^{+}\left( {qy} \right)\right]
\right|_{q=\beta_i}=
\left.2\imath Res \left[{h_l^{+} \left( {qx} \right) \frac{f_1
\left( q \right)}{f_2 \left( q \right) + \imath f_1 \left( q
\right)} } h_l^{+}\left( {qy} \right)\right]
\right|_{q=\beta_i}=\\
=2\imath \frac{f_1 \left( \beta_{i} \right)}{f'_2 \left( \beta_{i}
\right) + \imath f'_1 \left( \beta_{i} \right)} h_l^{+}\left(
{\beta_{i}x} \right)h_l^{+}\left( {\beta_{i}y} \right)=b_{i}
h_l^{+}\left( {\beta_{i}x} \right)h_l^{+}\left( {\beta_{i}y}
\right),
\end{array}
\end{equation}
here we have denoted $f'_{i}(q)=df_{i}(q)/dq$, ($i=1,\, 2$).


In this case the input kernel of eq. (\ref{eq2}) is a degenerate one as well as its output kernel
\begin{equation}
\label{eq8} L\left( {x,y} \right) = \sum\limits_{i = 1}^n {P_i
\left( x \right)h_l^ + \left( {\beta _i y} \right)} ,
\end{equation}

\noindent where $P_i \left( x \right)$ are unknown coefficients.
Substitution of (\ref{eq7}) and (\ref{eq8}) into (\ref{eq2})
yields
\begin{equation}
\label{eq2qa} \sum\limits_{i = 1}^n h_l^{+}\left({\beta_i y}\right) \left( { b_i h_l^{+} \left( {\beta_i x} \right) +
P_{i}(x) + b_{i}\sum\limits_{k = 1}^n {P_k \left( x \right) \int\limits_x^{ + \infty } {h_l^{+} \left( {\beta _{k} t}
\right) h_l^{+} \left( {\beta_i t} \right)}dt} }\right) = 0. \label{Mar22}
\end{equation}
Linear independence of the $h_l^{+}\left({\beta_i y}\right)$
implies that

\begin{equation}
\label{eq2fac}  b_i h_l^{+} \left( {\beta _i x} \right) + P_{i}(x)
+ b_{i}\sum\limits_{k = 1}^n {P_{k} \left( x \right)
\int\limits_x^{ + \infty } {h_l^{+} \left( {\beta _{k} t} \right)
h_l^{+} \left( {\beta _i t} \right)}dt}  = 0, \label{Mar23}
\end{equation}
or
\begin{equation}
\label{eq9}\sum\limits_{k = 1}^n {A_{ik} } \left( x \right)P_k
\left( x \right) = D_i\left( x \right)\ \ \ \ \ \ (i = 1,..,n),
\end{equation}

\noindent where $D_i \left( x \right) = - b_i h_l^ + \left( {\beta _i x} \right)$ and after applying Riccati-Hankel
integration formulas (see Appendix) in (\ref{eq2fac}) we have
\begin{equation} \label{eq101} \begin{array}{cc}
 A_{ik} =
 & \left\{
 \begin{array}{c}
 1+b_{i}x(\left(h_l^{+}\left( {\beta _i x}
\right)\right)^{2}-h_{l-1}^{+} \left( {\beta _i x}
\right)h_{l+1}^{+} \left( {\beta _i x} \right))/2,\ \ for\ \ i=k \\
-b_i \frac{\beta _i h_{l - 1}^{+} \left( {\beta _i x} \right)h_l^
+ \left( {\beta _k x} \right) - \beta _k h_l^{+} \left( {\beta _i
x} \right)h_{l - 1}^ + \left( {\beta _k x} \right)}{\beta _i^2 -
\beta _k^2},\ \ for\ \ i\neq k.
\end{array}\right.
 \end{array}
\end{equation}
The functional coefficients $P_i \left( x \right)$ are defined by
 (\ref{eq9})

\begin{equation}
\label{eq9a} P_i \left( x \right) = \left( {A^{ - 1}D} \right)_i.
\label{final_Mar}
\end{equation}

\noindent Finally we derive $L\left( {x,y} \right)$ and the potential $V\left( r \right)$ from (\ref{eq8}) and
(\ref{eq4}).

 In case of two coupled channels we
present only sketchy derivations because of their awkwardness. In this case the system of the partial Schr\"{o}dinger
equations is

\begin{equation}
\label{eq10} \left( {\frac{d^2}{dr^2} + V\left( r \right) + \left(
{{\begin{array}{*{20}c}
 {\frac{l_1 \left( {l_1 + 1} \right)}{r^2}} \hfill & 0 \hfill \\
 0 \hfill & {\frac{l_2 \left( {l_2 + 1} \right)}{r^2}} \hfill \\
\end{array} }} \right)} \right)\left( {{\begin{array}{*{20}c}
 {\chi _1 (r) } \hfill \\
 {\chi _2 (r)} \hfill \\
\end{array} }} \right) = q^2\left( {{\begin{array}{*{20}c}
 {\chi _1 (r)} \hfill \\
 {\chi _2 (r)} \hfill \\
\end{array} }} \right),
\end{equation}

\begin{equation}
\label{eq11} V\left( r \right) = \left( {{\begin{array}{*{20}c}
 {V_1 \left( r \right)} \hfill & {V_T \left( r \right)} \hfill \\
 {V_T \left( r \right)} \hfill & {V_2 \left( r \right)} \hfill \\
\end{array} }} \right) \quad ,
\end{equation}

\noindent where $V_1 \left( r \right)$, $V_2 \left( r \right)$ are
potentials in channels 1 and 2, $V_T \left( r \right)$ is
potential coupling them, $\chi _1 (r)$ and $\chi _2 (r)$ are
channel wave functions.

By analogy with  (\ref{eq5a}) we approximate the $S$-matrix by the following expression

\begin{equation}
\label{eqS2}
\begin{array}{l}
 S(x) = \left( {{\begin{array}{*{20}c}
 {\exp \left( {2\imath \delta _1 } \right)\cos 2\varepsilon } \hfill & {i\exp
\left( {\imath \left( {\delta _1 + \delta _2 } \right)}
\right)\sin 2\varepsilon }
\hfill \\
 {\imath \exp \left( {\imath \left( {\delta _1 + \delta _2 } \right)} \right)\sin
2\varepsilon } \hfill & {\exp \left( {2\imath \delta _2 }
\right)\cos 2\varepsilon
} \hfill \\
\end{array} }} \right) = \\
\hfill \\
 = \left( {{\begin{array}{*{20}c}
 {\left( {\frac{f_2^{\left( 1 \right)} \left( q \right) - \imath f_1^{\left( 1
\right)} \left( q \right)}{f_2^{\left( 1 \right)} \left( q \right)
+ \imath  f_1^{\left( 1 \right)} \left( q \right)}}
\right)^2\frac{\left( {f_2^{\left( {12} \right)} \left( q \right)}
\right)^2 - \left( {f_1^{\left( {12} \right)} \left( q \right)}
\right)^2}{\left( {f_2^{\left( {12} \right)} \left( q \right)}
\right)^2 + \left( {f_1^{\left( {12} \right)} \left( q \right)}
\right)^2}} \hfill
& { - 2 \imath  \frac{f_2^{\left( {12} \right)} \left( x
\right)f_1^{\left( {12} \right)} \left( x \right)}{\left(
{f_2^{\left( {12} \right)} \left( q \right)} \right)^2 + \left(
{f_1^{\left( {12} \right)} \left( q \right)}
\right)^2}\prod\limits_{j = 1,2} {\frac{f_2^{\left( j \right)}
\left( q \right) - \imath f_1^{\left( j \right)} \left( q
\right)}{f_2^{\left( j \right)} \left( q \right) + \imath
f_1^{\left( j
\right)} \left( q \right)}} } \hfill \\
 { - 2i \frac{f_2^{\left( {12} \right)} \left( q \right)f_1^{\left( {12}
\right)} \left( q \right)}{\left( {f_2^{\left( {12} \right)} \left( q \right)} \right)^2 + \left( {f_1^{\left( {12}
\right)} \left( q \right)} \right)^2}\prod\limits_{j = 1,2} {\frac{f_2^{\left( j \right)} \left( q \right) - i
f_1^{\left( j \right)} \left( q \right)}{f_2^{\left( j \right)} \left( q \right) + i f_1^{\left( j \right)} \left( q
\right)}} }  & {\left( {\frac{f_2^{\left( 2 \right)} \left( q \right) - \imath f_1^{\left( 2 \right)} \left( q
\right)}{f_2^{\left( 2 \right)} \left( q \right) + \imath  f_1^{\left( 2 \right)} \left( q \right)}}
\right)^2\frac{\left( {f_2^{\left( {12} \right)} \left( q \right)} \right)^2 - \left( {f_1^{\left( {12} \right)} \left(
q \right)} \right)^2}{\left( {f_2^{\left( {12} \right)} \left( q \right)} \right)^2 + \left( {f_1^{\left( {12} \right)}
\left( q
\right)} \right)^2}} \hfill \\
\end{array} }} \right) \\
 \end{array}
\end{equation}

This is again the most general Pad\'{e} approximant for the $S$-matrix. It was used in \cite{Wiesner} in an other form,
but corresponding analytical solution of the inverse scattering problem was not presented.

The coefficients of this Pad\'{e} approximant are determined from the equations analogous to (\ref{eq6b})
\begin{equation}
\label{eq6b} tg\left( { - \frac{\delta_i\left( q \right)}{2} } \right) = \frac{f^{(i)}_1 \left( q \right)}{f^{(i)}_2
\left( q \right)}, \ \ i=1,2
\end{equation}
and
\begin{equation}
\label{eq13} tg\left( {\varepsilon \left( q \right)} \right) = \frac{f_1 \left( q \right)}{f_2 \left( q \right)}.
\end{equation}

The generalized Marchenko equation for coupled channels formally has the former view \cite{Blaz}

\begin{equation}
\label{eq14} L\left( {x,y} \right) + F\left( {x,y} \right) +
\int\limits_x^{ + \infty } {L\left( {x,t} \right)F\left( {t,y}
\right)dt} = 0,
\end{equation}

\noindent but functions involved are matrices $\left( {2\times 2}
\right)$

\begin{equation}
\label{F2} F\left( {x,y} \right) = \frac{1}{2\pi }\int\limits_{ - \infty }^{ + \infty } {H\left( {qx} \right)\left[ {I
- S\left( q \right)} \right]H\left( {qy} \right)dq} + \sum\limits_{i = 1}^{n_{b} } {H\left( {\beta _i x} \right)M_i
H\left( {\beta _i y} \right)} ,
\end{equation}

\noindent where

\begin{equation}
\label{eq16} H\left( x \right) = \left( {{\begin{array}{*{20}c}
 {h_{l_1 }^ + \left( x \right)} \hfill & 0 \hfill \\
 0 \hfill & {h_{l_2 }^ + \left( x \right)} \hfill \\
\end{array} }} \right),\ I = \left( {{\begin{array}{*{20}c}
 1 \hfill & 0 \hfill \\
 0 \hfill & 1 \hfill \\
\end{array} }} \right).
\end{equation}

Insertion of (\ref{eqS2}) into (\ref{F2}) and applying of the
residue theorem yields

\begin{equation}
\label{eq17}
\begin{array}{l}
 F\left( {x,y} \right) = \imath\sum\limits_{i=1}^{n_{pos}}
\left.Res \left[H\left(qx\right)\left( {I - S\left( q \right)}
\right)H\left( {qy} \right)\right]\right|_{q=\beta_{i}} +
\sum\limits_{i = 1}^{n_{\mbox{\tiny{b}}} } {H\left( {\beta _i x}
\right)M^{2}_i H\left(
{\beta _i y} \right)} = \\
 = \sum\limits_{i=1}^{n} {H\left( {\beta _i x} \right)Q_i^1 H\left(
{\beta _i y} \right)} + \sum\limits_{i=1}^{n_{pos}^{(2)}} {x{H}'\left( {\beta _i x} \right)Q_i^2 H\left( {\beta _i y}
\right)}
 + \sum\limits_{i=1}^{n_{pos}^{(2)}}{H\left( {\beta _i x} \right)Q_i^2
{H}'\left( {\beta _i y} \right)y} , \\
 \end{array}
\end{equation}

\noindent where $\beta _i $ ($i = 1,...,n_{\mbox{\tiny{pos}}} )$ are all $S$-matrix poles with $\Im\, \beta_i>0$,
$\beta _i $ ($i = 1,...,n_{\mbox{\tiny{pos}}}^{\tiny{(2)}} )$ are poles of the second order,
$\beta=\{\beta_{1},..\beta_{n_{\mbox{\tiny{pos}}}^{\tiny{(2)}}},..
\beta_{n_{pos}},\tilde{q_{1}},...\tilde{q_{n_{b}}}\}$, $n=n_{pos}+n_{b}$,
\[
{H}'\left( x \right) = \left( {{\begin{array}{*{20}c}
 {{dh_{l_1 }^ + \left( x \right)} \mathord{\left/ {\vphantom {{dh_{l_1 }^ +
\left( x \right)} {dx}}} \right. \kern-\nulldelimiterspace} {dx}}
\hfill & 0
\hfill \\
 0 \hfill & {{dh_{l_2 }^ + \left( x \right)} \mathord{\left/ {\vphantom
{{dh_{l_2 }^ + \left( x \right)} {dx}}} \right.
\kern-\nulldelimiterspace}
{dx}} \hfill \\
\end{array} }} \right).
\]
We note that there are poles of the first as well as of the second order in the diagonal matrix elements and there are
poles only of the first order in  the off-diagonal matrix elements. Poles of the second order in the diagonal elements
are poles of the first order  in  the off-diagonal matrix elements, and they must be enumerated twice. $Q_i^j\ \ \left(
{j = 1,2} \right)$ are constant matrices they are trivial but cumbersome therefore, we do not give them.

We solve eq. (\ref{eq14}) using substitution

\begin{equation}
\label{eq18} L\left( {x,y} \right) = \sum\limits_{i = 1}^n {P_i
\left( x \right)} H\left( {\beta _i y} \right) + \sum\limits_{i =
1}^n {N_i \left( x \right)} y{H}'\left( {\beta _i y} \right),
\end{equation}
where $P_i \left( x \right)$, $N_i \left( x \right)$ are unknown
functional $\left( {2\times 2} \right)$ matrix-coefficients.
Linear independence of the $H\left( {\beta _i y} \right)$ and
$y{H}'\left( {\beta _i y} \right)$ implies that

\begin{equation}
\label{sys2}\begin{array}{r}
 \sum\limits_i {P_i \left( x \right)Q_{ij}^3 \left( x \right)} +
\sum\limits_i {N_i \left( x \right)} Q_{ij}^5 \left( x \right) =
H\left(
{\beta _j x} \right)Q_j^1 + x{H}'\left( {\beta _j x} \right)Q_j^2 \\
 \sum\limits_i {N_i \left( x \right)Q_{ij}^6 \left( x \right)} +
\sum\limits_i {P_i \left( x \right)} Q_{ij}^4 \left( x \right) =
H\left(
{\beta _j x} \right)Q_j^2 \\
 \end{array}\end{equation}

\noindent where

\[
Q_{ij}^3 \left( x \right) = I\delta _{ij} + \int\limits_x^{ +
\infty } {H\left( {\beta _i t} \right)H\left( {\beta _j t}
\right)dt\times Q_j^1 } + \int\limits_x^{ + \infty } {tH\left(
{\beta _i t} \right){H}'\left( {\beta _j t} \right)dt\times Q_j^2
}
\]

\begin{equation}
\label{eq19} Q_{ij}^4 \left( x \right) = \int\limits_x^{ + \infty
} {H\left( {\beta _i t} \right)H\left( {\beta _j t}
\right)dt\times Q_j^2 }
\end{equation}

\[
Q_{ij}^5 \left( x \right) = \int\limits_x^{ + \infty }
{t{H}'\left( {\beta _i t} \right)H\left( {\beta _j t}
\right)dt\times Q_j^1 } + \int\limits_x^{ + \infty }
{t^2{H}'\left( {\beta _i t} \right){H}'\left( {\beta _j t}
\right)dt\times Q_j^2 }
\]

\[
Q_{ij}^6 \left( x \right) = I\delta _{ij} + \int\limits_x^{ +
\infty } {t{H}'\left( {\beta _i t} \right)H\left( {\beta _j t}
\right)dt\times Q_j^2 } ,
\]

\noindent Integrals of expressions (\ref{eq19}) are presented in Appendix. Matrix equations (\ref{sys2}) can be
trivially reduced to scalar linear equations. Having solved this linear equation system we get the sought-for potential
from (\ref{eq18}) and (\ref{eq4}).

The multichannel generalization is trivial.

\section{The optical potential}
\mbox{}

In this section we consider changes of $S$-matrix that are induced by certain transformation of real potential.

First we consider the one channel problem.

The phase equation \cite{Calo} for the initial potential
$V^0\left( r \right)$ obtained by some inversion procedure (from
Marchenko equation in our calculations) is

\begin{equation}
\label{eq20} \delta _l^{(0)} = - \frac{1}{q}\int\limits_0^\infty
{V^{(0)}\left( r \right)\hat{D}_l^2 \left( {qr} \right)\sin
^2\left( {\hat{\delta}_{l}(qr) + \delta ^{(0)}\left( r \right)}
\right)dr} ,
\end{equation}

\noindent where  $\hat{D}_l \left( z \right)$ and  $\hat{\delta}_l \left( z \right)$ are Riccati-Bessel amplitude and
phase correspondingly \cite{Calo}
\begin{equation}
\label{eqDD} \hat{D}_{l}(x)=\sqrt{j_l^2(x)+n_l^2(x)},
\end{equation}
\begin{equation}
\label{eqdd} \hat{\delta}_l(x)=-\arctan(j_l(x)/n_l(x)) \label{qr}
\end{equation}

Let us consider the complex-valued potential $V^{(1)}\left( r
\right)$ obtained from $V^0\left( r \right)$ by transformation

\begin{equation}
\label{eq21} V^{(1)}\left( r \right) = \left( {1 + i\alpha }
\right)V^{(0)}\left( r \right),
\end{equation}

\noindent where $\alpha $ is some real parameter. Such parametrization was used in \cite{Neud2} but without analysis
($\alpha$ was fitted). Evidently the phase equation for this potential is

\begin{equation}
\label{eq22} \delta ^{(1)} = - \frac{1}{q}\left( {1 + i\alpha }
\right)\int\limits_0^\infty {V^{(0)}\left( r \right)\hat{D}_l^2
\left( {qr} \right)\sin ^2\left( {\hat{\delta}_{l}(qr) + \delta
^{(1)}\left( r \right)} \right)dr} .
\end{equation}

From eqs. (\ref{eq20}) and (\ref{eq22}) we get

\begin{equation}
\label{eq23}
\begin{array}{l}
 \delta ^{(1)} - \left( {1 + i\alpha } \right)\delta ^{(0)} = \\
 = - \frac{1 + i\alpha }{q} \int\limits_0^\infty
{V^{(0)}\left( r \right)\hat{D}_l^2 \left( {qr} \right)\left(
{\sin ^2\left( {\hat{\delta}_{l}(qr) + \delta ^{(1)}\left( r
\right)} \right) - \sin ^2\left( {\hat{\delta}_{l}(qr) + \delta
^{(0)}\left( r \right)} \right)} \right)dr} = \label{difference}\\
 = - \frac{1 + i\alpha }{q}\int\limits_0^\infty
{V^{(0)}\left( r \right)\hat{D}_l^2 \left( {qr} \right)\underline{\sin \left( {2\hat{\delta}_{l}(qr) + \delta
^{(1)}\left( r \right) + \delta ^{(0)}\left( r \right)} \right)}\sin \left( {\delta ^{(1)}\left( r \right) - \delta
^{(0)}\left( r \right)} \right)dr}
 \end{array}
\end{equation}

For smooth enough potentials  the right side of eq. (\ref{eq23}) rapidly decreases comparing with
 $\delta ^{(0)}$ and $\delta ^{(1)}$, because
 there is a rapidly oscillating around zero function under the integral in (\ref{eq23}) (underlined).
 Its frequency behaves as $2q$ for big $q$ (see (\ref{qr})).
 Then as the first approximation we may take

\begin{equation}
\label{eq24} \delta ^{(1)} \approx \left( {1 + i\alpha }
\right)\delta ^{(0)} = \delta _R + i\delta _I .
\end{equation}

For inelastic scattering the $S$-matrix is expressed through the real inelasticity parameter  $\rho $ and the real
phase shift $\delta$

\begin{equation}
\label{eq25} S = \cos ^2\left( \rho \right)e^{2i\delta } =
e^{2i\left( {\delta _R + i\delta _I } \right)},
\end{equation}

\noindent so we easily arrive at

\begin{equation}
\label{eq26} \delta_R=\delta \approx \delta ^{(0)},
\end{equation}

\begin{equation}
\label{eq27} \cos ^2\left( \rho \right) \approx e^{ - 2\alpha
\delta ^{(0)}},
\end{equation}

\noindent whence it follows that $\alpha \delta \ge 0$. The
formula (\ref{eq27}) allows to calculate the parameter $\alpha $
from the known values $\rho $ and $\delta ^{(0)} \approx \delta $.

Eqs. (\ref{difference}-\ref{eq24}) imply that
\begin{equation}
\label{imply} \int\limits_0^\infty {V^{(0)}\left( r \right)\hat{D}_l^2 \left( {qr} \right)\sin ^2\left(
{\hat{\delta}_{l}(qr) + \delta ^{(0)}\left( r \right)} \right)dr}\simeq  \int\limits_0^\infty {V^{(0)}\left( r
\right)\hat{D}_l^2 \left( {qr} \right)\sin ^2\left( {\hat{\delta}_{l}(qr) + \delta ^{(1)}\left( r \right)} \right)dr},
\end{equation}

%
%

Consideration of the coupled partial waves is more complicated.
The initial real potential is
\begin{equation}
\label{eq28b} V^{(0)}\left( r \right) =\left( {{\begin{array}{*{20}c}
 {V_{1}^{(0)} } \hfill & {V_{T}^{(0)} } \hfill \\
{V_{T}^{(0)} } & {V_{2}^{(0)}
} \hfill \\
\end{array} }} \right).
\end{equation}
%
The equations for eigenphases and mixing parameter of potential (\ref{eq28b}) are \cite{Calo}

\begin{eqnarray}
\delta_{1}^{(1)} =  I^{(0)}_{11}+ I^{(0)}_{12}+I^{(0)}_{13}\label{dfromI1}\\
\delta_{2}^{(1)} =  I^{(0)}_{21}+ I^{(0)}_{22}+I^{(0)}_{23}\\
\epsilon^{(1)} =   I^{(0)}_{31}+I^{(0)}_{32}+I^{(0)}_{33},\label{dfromI3}
\end{eqnarray}
where

\begin{eqnarray}
I^{(0)}_{11} =-\frac{1}{q}\int\limits_0^\infty dr
V_{1}^{(0)}(r)\cos^{2}\epsilon^{(0)}(r)\hat{D}_{l_{1}}^{2}(qr)\sin^2(\hat{\delta}_{l_{1}}(qr)+\delta_{1}^{(0)}(r))
\nonumber\\
I^{(0)}_{12} =-\frac{1}{q}\int\limits_0^\infty dr V_{2}^{(0)}(r)\sin^{2}\epsilon^{(0)}(r)\hat{D}_{l_{2}}^{2}(qr)
\sin^2(\hat{\delta}_{l_{2}}(qr)+\delta_{1}^{(0)}(r))\nonumber\\
I^{(0)}_{13} =-\frac{1}{q}\int\limits_0^\infty dr V_{T}^{(0)}(r)\sin
2\epsilon^{(0)}(r)\hat{D}_{l_{2}}(qr)\sin(\hat{\delta}_{l_{2}}(qr)+\delta_{1}^{(0)}(r))
\hat{D}_{l_{1}}(qr)\sin(\hat{\delta}_{l_{1}}(qr)+\delta_{1}^{(0)}(r))\label{eqI1}
\end{eqnarray}

\begin{eqnarray}
I^{(0)}_{21} =-\frac{1}{q}\int\limits_0^\infty dr
V_{1}^{(0)}(r)\sin^{2}\epsilon^{(0)}(r)\hat{D}_{l_{1}}^{2}(qr)\sin^2(\hat{\delta}_{l_{1}}(qr)+\delta_{2}^{(0)}(r))
\nonumber\\
I^{(0)}_{22} =-\frac{1}{q}\int\limits_0^\infty dr V_{2}^{(0)}(r)\cos^{2}\epsilon^{(0)}(r)
\hat{D}_{l_{2}}^{2}(qr)\sin^2(\hat{\delta}_{l_{2}}(qr)+\delta_{2}^{(0)}(r))\nonumber\\
I^{(0)}_{23} =-\frac{1}{q}\int\limits_0^\infty dr V_{T}^{(0)}(r)\sin
2\epsilon^{(0)}(r)\hat{D}_{l_{2}}(qr)\sin(\hat{\delta}_{l_{2}}(qr)+\delta_{2}^{(0)}(r))
\hat{D}_{l_{1}}(qr)\sin(\hat{\delta}_{l_{1}}(qr)+\delta_{2}^{(0)}(r))\label{eqI2}
\end{eqnarray}
\begin{eqnarray}
I^{(0)}_{31} =\frac{1}{2q}\int\limits_0^\infty
\frac{\sin2\epsilon^{(0)}(r)dr}{\sin(\delta_{1}^{(0)}(r)-\delta_{2}^{(0)}(r))}
V_{1}^{(0)}(r)\hat{D}^{2}_{l_{1}}(qr)\sin(\hat{\delta}_{l_{1}}(qr)+\delta_{1}^{(0)}(r))
\sin(\hat{\delta}_{l_{1}}(qr)+\delta_{2}^{(0)}(r))\nonumber\\
I^{(0)}_{31} =-\frac{1}{2q}\int\limits_0^\infty
\frac{\sin2\epsilon^{(0)}(r)dr}{\sin(\delta_{1}^{(0)}(r)-\delta_{2}^{(0)}(r))}
V_{2}^{(0)}(r)\hat{D}^{2}_{l_{2}}(qr)\sin(\hat{\delta}_{l_{2}}(qr)+\delta_{1}^{(0)}(r))
\sin(\hat{\delta}_{l_{2}}(qr)+\delta_{1}^{(0)}(r))\nonumber\\
I^{(0)}_{31} =-\frac{1}{2q}\int\limits_0^\infty
\frac{V_{T}^{(0)}(r)\hat{D}_{l_{1}}(qr)\hat{D}_{l_{2}}(qr)dr}{\sin(\delta_{1}^{(0)}(r)-\delta_{2}^{(0)}(r))}
\left[\cos2\epsilon^{(0)}(r)\sin(\hat{\delta}_{l_{1}}(qr)+
\delta_{1}^{(0)}(r))\sin(\hat{\delta}_{l_{2}}(qr)+\delta_{2}^{(0)}(r))-\right.\nonumber\\
\left.-\frac{1}{2}\left(\cos2\epsilon^{(0)}(r)-1\right)\sin(\delta_{1}^{(0)}(r)-\delta_{2}^{(0)}(r))
\sin\left(\hat{\delta}_{l_{1}}(qr)-\hat{\delta}_{l_{2}}(qr)\right)\right]\label{eqE}
\end{eqnarray}

By analogy with the one channel case the following generalization
for the optical potential is derived

\begin{equation}
\label{eq28} V^{(1)}\left( r \right) =\left( {{\begin{array}{*{20}c}
 {\left( {1 + i\alpha_{1} } \right)V_{1}^{(0)} } \hfill & {\left( 1 + i
\alpha_{3}
\right)V_{T}^{(0)} } \hfill \\
{\left( 1 + i \alpha_{3} \right)V_{T}^{(0)} } & {\left( {1 + i\alpha_{2} } \right)V_{2}^{(0)}
} \hfill \\
\end{array} }} \right).
\end{equation}

Evidently the phase equations for this potential is
\begin{eqnarray}
\delta_{1}^{(1)} = {\left( {1 + i\alpha_{1} }\right)} I^{(1)}_{11}+{\left( {1 + i\alpha_{2} }\right)}
I^{(1)}_{12}+{\left( {1 +
i\alpha_{3} }\right)}I^{(1)}_{13}\label{eqNewDD1} \\
\delta_{2}^{(1)} = {\left( {1 + i\alpha_{1} }\right)} I^{(1)}_{21}+{\left( {1 + i\alpha_{2} }\right)}
I^{(1)}_{22}+{\left( {1 +
i\alpha_{3} }\right)}I^{(1)}_{23}\\
\epsilon^{(1)} =  {\left( {1 + i\alpha_{1} }\right)} I^{(1)}_{31}+{\left( {1 + i\alpha_{2} }\right)}
I^{(1)}_{32}+{\left( {1 + i\alpha_{3} }\right)}I^{(1)}_{33} \label{eqNewDD}
\end{eqnarray}

Integrals $I^{(1)}_{ij}$ are defined as $I^{(0)}_{ij}$ in (\ref{eqI1}-\ref{eqE}) but through $\delta_{1}^{(1)}(r)$,
$\delta_{2}^{(1)}(r)$ and $\epsilon^{(1)}(r)$ instead of  $\delta_{1}^{(0)}(r)$, $\delta_{2}^{(0)}(r)$ and
$\epsilon^{(0)}(r)$.

Evidently we cannot consider (\ref{eqNewDD1}-\ref{eqNewDD}) in a manner like (\ref{eq23}). But we assume that
\begin{equation}
I^{(1)}_{ij}= I^{(0)}_{ij}+\sum\limits_{i,j=1,2,3} o(I^{(0)}_{ij}),\label{approxx}
\end{equation}
where
\begin{equation}
\sum\limits_{i,j=1,2,3} o(I^{(0)}_{ij})\ll I^{(0)}_{ij},\ \ \mbox{ for } i,j=1,2,3.
\end{equation}

 This assumption can be considered as a generalization of
 (\ref{imply}). It is hard to prove in the general case, but our calculations show that this is true at least
 in case of $^{3}SD_1$ NN scattering.

Eigenphases $\hat{\delta}^{(0)}_{i}$, $i=1,2$ and mixing parameter $\hat{\epsilon}^{(0)}$ are real and they define a
unitary $S^{(0)}$-matrix
\begin{equation}
\label{EigenS2} S^{(0)}=\left( \begin {array}{cc}  \cos^2
\hat{\epsilon}^{(0)} {e^{2\,\imath{\hat \delta_1}^{(0)}}}+ \sin^2
\hat{\epsilon}^{(0)} {e^{2\,\imath{\hat \delta_2}^{(0)}}}&
\cos \hat{\epsilon}^{(0)}\sin \hat{\epsilon}^{(0)}\left(
  {e^{2\,\imath{\hat \delta_1}^{(0)}}}-{e^{2\,i{\hat \delta_2}^{(0)}}}
 \right) \\
 \noalign{\medskip}\cos \hat{\epsilon}^{(0)}\sin \hat{\epsilon}^{(0)}\left(
  {e^{2\,\imath{\hat \delta_1}^{(0)}}}-{e^{2\,i{\hat \delta_2}^{(0)}}}
 \right) &
 \sin^2  \hat{\epsilon}^{(0)} {e^{2\,\imath{\hat \delta_1}^{(0)}}}+
\cos^2 \hat{\epsilon}^{(0)} {e^{2\,\imath{\hat \delta_2}^{(0)}}}\end {array} \right)
\end{equation}

 Eigenphases
$\hat{\delta}^{(1)}_{i}$ and mixing parameter $\hat{\epsilon}^{(1)}$ are complex but they define $S^{(1)}$-matrix in
the regular way
\begin{equation}
\label{EigenS} S^{(1)}=\left( \begin {array}{cc}  \cos^2  \hat{\epsilon}^{(1)} {e^{2\,\imath{\hat \delta_1}^{(1)}}}+
\sin^2 \hat{\epsilon}^{(1)} {e^{2\,\imath{\hat \delta_2}^{(1)}}}&
\cos \hat{\epsilon}^{(1)}\sin \hat{\epsilon}^{(1)}\left(
  {e^{2\,\imath{\hat \delta_1}^{(1)}}}-{e^{2\,i{\hat \delta_2}^{(1)}}}
 \right) \\
 \noalign{\medskip}\cos \hat{\epsilon}^{(1)}\sin \hat{\epsilon}^{(1)}\left(
  {e^{2\,\imath{\hat \delta_1}^{(1)}}}-{e^{2\,i{\hat \delta_2}^{(1)}}}
 \right) &
 \sin^2  \hat{\epsilon}^{(1)} {e^{2\,\imath{\hat \delta_1}^{(1)}}}+
\cos^2 \hat{\epsilon}^{(1)} {e^{2\,\imath{\hat \delta_2}^{(1)}}}\end {array} \right).
\end{equation}
The complex  eigenphases $\hat{\delta}^{(1)}_{i}$ and mixing parameter $\hat{\epsilon}^{(1)}$ are defined by the
experimental $S$-matrix ($S\equiv S^{(1)}$). A direct consequence of (\ref{eqNewDD1}-\ref{approxx}) is
\begin{equation}
\Re\delta_{i}^{(1)}=\delta_{i}^{(0)},\ \ i=1,2;\ \ \Re\epsilon^{(1)}=\epsilon^{(0)}.\label{CouplDelta}
\end{equation}
\begin{equation}
\Im\delta_{i}^{(1)}=\sum\limits_{j=1}^{3}\alpha_{j}I^{(0)}_{ij},\ \ i=1,2;\ \
\Im\epsilon_{i}^{(1)}=\sum\limits_{j=1}^{3}\alpha_{j}I^{(0)}_{3j}.\label{CouplDelta2}
\end{equation}
We may calculate coefficients $I^{(1)}_{ij}\simeq I^{(0)}_{ij}\ (i,j=1,2,3)$ using (\ref{eqI1}-\ref{eqE}). A simpler
method is to use the following implication of (\ref{eqNewDD1}-\ref{approxx})
\begin{equation}
I^{(1)}_{ij}=\Im\frac{ \partial \delta_{i}^{(1)} }{\partial \alpha_{j}},\ i=1,2;\ \ I^{(1)}_{3j}=\Im\frac{ \partial
\epsilon^{(1)} }{\partial \alpha_{j}}\label{Simpler}.
\end{equation}
Calculated coefficients may be checked by (\ref{dfromI1}-\ref{dfromI3}). Next, we calculate $\alpha_{i}\ (i=1,2,3)$
from (\ref{eqNewDD1}-\ref{eqNewDD}).

\section{The optical potentials }
  \mbox{}
We apply the developed method of inversion to analysis  $NN$, $\pi^{-} N$  and $K^{+}N$ data up to energies where
relativistic effects are essential. We take into account these effects in the frames of relativistic quantum mechanics
of systems with a fixed number of particles. The review of this approach can be found in \cite{RQM}. Here we give only
some extracts of it.

The relativistic quantum mechanics of systems consisting of a fixed number of particles is based on the conjecture that
the number of particles is constant at not very high energies and on the assumption that the group of invariance for
the system under consideration is  the Poincare group rather than the Galilei one.  A system of two particles is
described by the wave function, which is an eigenfunction of the mass operator. In this case we may represent this wave
function as a product of the external and internal wave functions \cite{Lev,Khokhlov}. The internal wave function
$\chi$ is also an eigenfunction of the mass operator and satisfies the following equation
\begin{equation}
\label{rel1} \left[ \sqrt{\hat{q}^2+m_1^2}+\sqrt{\hat{q}^2+m_2^2}+V_{int} \right]\chi = M\chi,
\end{equation}
where $V_{int}$ is some interaction operator acting only through internal variables (spins and relative momentum),
$\hat{q}$ is a momentum operator of one of the particles in the center of masses frame. Rearrangement of (\ref{rel1})
gives
\begin{equation}
\label{rel2} \left[ {\hat{q}^2 + 2m V} \right]\chi = q^2\chi,\label{eq56}
\end{equation}
where
\begin{equation}
\label{rel3} q^2=\frac{M^2}{4}-\frac{m_1^2+m_2^2}{2}+\frac{(m_1^2-m_2^2)^2}{4M^2},
\end{equation}
$m$ is taken as a nonrelativistic reduced mass
\begin{equation}
\label{reducedM} m=\frac{m_1 m_2}{m_1+m_2},
\end{equation}
$V$ is an operator acting like $V_{int}$ only through internal variables.

In case of two particles with equal masses $m_{1}=m_{2}\equiv 2m$
\begin{equation}
  q^2 = \frac{M^2}{4} -
4m^2.
\end{equation}

Eq. (\ref{eq56}) is identical in form to the Schr\^{o}dinger equation.
The quasicoordinate representation corresponds to the realization  ${\rm {\bf q}} = - i\frac{\partial }{\partial {\rm
{\bf r}}}$, $V=V({\rm {\bf r}})$.  In \cite{Khokhlov2} we showed that this formalism can be easily generalized for the
case of inelastic channels, particularly it allows to take into account isobar channels in NN scattering.
 This formal coincidence allows us to apply our inversion algorithm.

 We applied  the described algorithm of inversion to reconstruction
 of the nucleon-nucleon potential. As input data for this reconstruction
 we used modern phase shift analysis data  up to  1100 MeV for ${
}^3SD_1 $ state and up to 3 GeV for ${ }^1S_0 $ state of nucleon-nucleon system \cite{DataScat,cite}. The deuteron
properties were taken from \cite{DataDeut}. These data allow to construct nucleon-nucleon partial potentials sustaining
forbidden bound states (Moscow potential introduced in \cite{Neud}). Parameters of forbidden bound states for the
partial Moscow potentials were chosen to be equal to those of model potentials (see Sect. 2). In this way we
constructed the NN optical potentials  for $^1 S_0$ and $^3 SD_1$ partial waves. These potentials describe the deuteron
properties and the phase shift analysis data. The $^1 S_0$  phase shifts of Moscow potential begin from $\pi$. $^3 S_0$
phase shifts of Moscow potential begin from $2\pi$. The mixing parameter $\epsilon_1$ of Moscow potential differs from
that of traditional repulsive core potential by sign. The real parts of the constructed partial potentials are
presented in fig. 1. The calculated values of deuteron properties are compared with the experimental data
\cite{DataDeut} in Table 1. Only three parameters are fixed as input data of inversion problem. These parameters are
energy, $A_S$ and $\eta_{d/s}$. Figs. 1 and 3 demonstrate how changes of $\delta$ influence the partial potential for
$^1 S_0$ wave.

 As
another example of application we analyzed the modern $P31$ $\pi^{-} N$ data up to 2 GeV and $S01$ $K^{+}N$ data up to
1 GeV \cite{cite} and constructed the corresponding optical potentials.   The real parts of the constructed partial
potentials for $P31$ $\pi^{-} N$ data up to 2 GeV and $S01$ $K^{+}N$ are presented in fig. 2.

%

From eqs. (\ref{eq27}) and (\ref{CouplDelta2}, \ref{Simpler}) we calculated parameters $\alpha$ and $\alpha_{i},\
(i=1,2,3)$ which define the imaginary parts of potentials. Our predictions were justified. Calculations with optical
and real potentials (fig. 3,7,8) show the validity of (\ref{eq26}) and (\ref{CouplDelta}). The $\alpha$'s predicted by
(\ref{eq27}) may be improved by a simple numerical method. Predicted and improved  values of $\alpha$'s are shown in
fig. 4,5,6. In all figures "Calc. I" means calculations from predicted values of (\ref{eq26}), "Calc. II" means
calculations from refined values. Fig. 5 shows that calculation of $\alpha_i,\ (i=1,2,3)$ from (\ref{CouplDelta2} and
\ref{Simpler}) does not require refinement because it uses more precise values of $I^{(1)}_{ij}$ than those implied by
assumption that $I^{(1)}_{ij}=I^{(0)}_{ij}$.

 In eq. (\ref{eq25}) we use parametrization of the $S$-matrix from \cite{Ray} whereas
 parametrization of partial wave analysis \cite{DataScat} is based on type-$K$ scheme \cite{SArndt,Bryan}.
 For uncoupled waves, the $S$-matrix is given by
\begin{equation}
\label{SArndt1}
\begin{array}{c}
S=\frac{1-K_i+\imath K_r}{1+K_i-\imath K_r},\\
\hfill\\ \mbox{where   }K_r=tan\tilde{\delta},\ K_i=tan^{2}\tilde{\rho}.
\end{array}
\end{equation}
So we had to recalculate data of \cite{DataScat} into $S$-matrix, then into parameters of \cite{Ray} to get input data
of inverse problem. Our results are presented in parametrization of \cite{DataScat}.
%


All potentials and
inelasticity multipliers ($\alpha$'s)  may be downloaded from cite www.physics.khstu.ru in numerical form.

\section{Conclusions}

Let us summarize the results presented in this work. In the first place we mention a presented analytical solution of
the Marchenko equation for coupled partial waves in case of diagonal Pad\'{e} approximant of the corresponding
$S$-matrix. The inverse scattering scheme at fixed angular momentum is used to construct a local real energy
independent potential as a first step of our inversion procedure for single and coupled waves. Furthermore, we consider
what changes of $S$-matrix are induced by certain transformation of real potential. We have found out that certain
simple transformation may have a negligible effect on phase shift but introduce a controllable inelasticity. This
transformation does not change the real part of the potential but adds an imaginary part. As a result we get an optical
potential with energy independent real part and energy dependent imaginary part.   We apply this scheme to NN, $\pi^{-}
N$ and $K^{+}N$ scattering successfully.

\section{Appendix}

We present only nontrivial integrals of (\ref{eq19}). They can be derived from the recursion relations for the
Riccati-Hankel functions and from known integrals \cite{Grad}.

\begin{equation} \label{A1} \begin{array}{cc}
 I_{1}(x,\beta_{i},\beta_{k},l)=\int\limits_x^\infty {h^{+}_{l}(\beta_{i}t)
h^{+}_{l}(\beta_{k})dt}=
 & \left\{
 \begin{array}{c}
 -x(\left(h_l^{+}\left( {\beta _i x}
\right)\right)^{2}-h_{l-1}^{+} \left( {\beta _i x}
\right)h_{l+1}^{+} \left( {\beta _i x} \right))/2,\ \ for\ \ i=k \\
\frac{\beta _i h_{l - 1}^{+} \left( {\beta _i x} \right)h_l^ +
\left( {\beta _k x} \right) - \beta _k h_l^{+} \left( {\beta _i x}
\right)h_{l - 1}^ + \left( {\beta _k x} \right)}{\beta _{ik}},\ \
for\ \ i\neq k.
\end{array}\right.
 \end{array}
\end{equation}
\begin{equation} \label{A2} \begin{array}{cc}
 \int\limits_x^\infty {h'^{+}_{l}(\beta_{i}t)
h^{+}_{l}(\beta_{k}t)tdt}=
 & \left\{
 \begin{array}{c}
-\left(x\left(h_l^{+}\left( {\beta _i x}
\right)\right)^{2}+I_{1}(x,\beta_{i},\beta_{i},l)\right)/(2\beta_{i}),\ \ for\ \ i=k \\
\begin{array}{c}
\frac{2\beta_{i}}{\beta_{ik}^2}\left(\beta_{k} h_l^{+}\left(
{\beta _i x} \right)h_{l-1}^{+}\left( {\beta _k x} \right)
-\beta_{i} h_{l-1}^{+}\left( {\beta _i x} \right)h_{l}^{+}\left(
{\beta _k x} \right)\right)+\\
+ \frac{1}{\beta_{ik}} h_{l-1}^{+}\left( {\beta _i x}
\right)h_{l}^{+}\left( {\beta _k x}\right) -\\
-\frac{x}{\beta_{ik}} \left(\beta_{k} h_l'^{+}\left( {\beta _i x}
\right)h_{l-1}^{+}\left( {\beta _k x} \right) -\beta_{i}
h_{l-1}'^{+}\left( {\beta _i x} \right)h_{l}^{+}\left( {\beta _k
x} \right)\right) ,\ \ for\ \ i\neq k.
\end{array}\end{array}\right.
 \end{array}
\end{equation}
Where $\beta_{ik}=\beta_{i}^2-\beta_{k}^2$

\begin{equation} \label{A4} \begin{array}{cc}
 \int\limits_x^\infty {h'^{+}_{l}(\beta_{i}t)
h'^{+}_{l}(\beta_{k}t)t^2dt}=
 & \left\{
 \begin{array}{c}
I_{2}\left( \beta_{i}x,l\right)/\beta_{i}^3,\ \ for\ \ i=k \\
\begin{array}{c}
8\frac{\beta_{i}\beta_{k}}{\beta_{ik}^3} \left(
\beta_{k}h_l^{+}\left( {\beta _i x}\right)h_{l-1}^{+}\left(
{\beta_k x}\right)-\beta_{i}h_l^{+}\left( {\beta _k
x}\right)h_{l-1}^{+}\left( {\beta_i x}\right)
\right)+\\
+2\frac{\beta_{i}}{\beta_{ik}^2} \left(
h_l^{+}\left( {\beta _i x}\right)h_{l-1}^{+}\left( {\beta_k x}\right)+\beta_{k}x h_l^{+}\left( {\beta _i
x}\right)h_{l-1}'^{+}\left( {\beta_k x}\right)-\right.\\
\left. -\beta_{i}x h_l'^{+}\left( {\beta _k x}\right)h_{l-1}^{+}\left( {\beta_i x}\right)
\right)+\\
%
%
%
+\frac{x}{\beta_{ik}} \left(
h_l'^{+}\left( {\beta _k x}\right)h_{l-1}^{+}\left( {\beta_i x}\right)-h_l'^{+}\left( {\beta _i
x}\right)h_{l-1}^{+}\left( {\beta_k x}\right)-\right.\\
\left. -\beta_{k}x h_l'^{+}\left( {\beta _i x}\right)h_{l-1}'^{+}\left( {\beta_k x}\right)
 +\beta_{i}x h_l'^{+}\left( {\beta _k
x}\right)h_{l-1}'^{+}\left( {\beta_i x}\right)
\right)-\\
+\frac{2\beta_{k}}{\beta_{ik}^2} \left(
h_l^{+}\left( {\beta_k x}\right)h_{l-1}^{+}\left( {\beta_i x}\right)-\beta_{k}xh_l'^{+}\left( {\beta _i
x}\right)h_{l-1}^{+}\left( {\beta_k x}\right)+\right.\\
\left.+\beta_{i}xh_l^{+}\left( {\beta _k x}\right)h_{l-1}'^{+}\left( {\beta_i x}\right) \right)
 ,\ \ for\ \ i\neq k.
\end{array}\end{array}\right.

 \end{array}
\end{equation}
\begin{equation}
I_{2}\left(z,1  \right)=\left(\imath
z^2/2-3z/2-9\imath/2+1/z\right)\exp(\imath 2z)
\end{equation}
\begin{equation}
I_{2}\left(z,2 \right)=\left(-\imath z^2/2+7z/2+49\imath/2-24/z-24\imath/x^2+12/x^3\right)\exp(\imath 2z)
\end{equation}
\begin{equation}
I_{2}\left(z,3  \right)=\left(\imath
z^2/2-13z/2-169\imath/2+171/z+450\imath/x^2-765/x^3-810\imath/x^4+405/x^5\right)\exp(\imath 2z)
\end{equation}
\begin{equation}
\begin{array}{c}
I_{2}\left(z,4  \right)=\left(-\imath z^2/2+21z/2+441\imath/2-745/z-3510\imath/x^2+11835/x^3+\right.\\
\left.+28560\imath/x^4-47880/x^5-50400\imath/x^6+25200/x^7 \right)\exp(\imath 2z)
\end{array}
\end{equation}


\newpage

\renewcommand{\baselinestretch}{1}
\begin{table}[t]
{Table 1. The deuteron properties}\\
\begin{tabular}
{|p{87pt}|p{87pt}|p{73pt}|} \hline & Exp. $^a$ & Calculation\\
&  & with Moscow \\
 &     & potential \\          \hline Energy (MeV) & 2,22458900(22)&  2,2246$^a$ \\
\hline $Q$ (Fm$^{2}$)& 0,2859(3)& 0,277$^c$\\
\hline A$_{S}$  (Fm$^{ - 1 / 2})$& 0,8802(20)&  0,8802\\
\hline r$_{d}$ (Fm)& 1,9627(38)&  1,956\\
\hline  $\eta_{d / s} $ & 0,02714 & 0,02714\\
\hline  $\mu _{d}$& 0,857406(1)&  0,859$^c$\\
\hline
\end{tabular}
\label{tab1}\vspace*{0.5cm}\\
$^a$ relativistic correction included; $^b$ Data are from
\cite{DataDeut}; $^c$ Meson exchange currents are not included.
\end{table}

\begin{figure}[t] \epsfysize=100mm \centerline{
\epsfbox{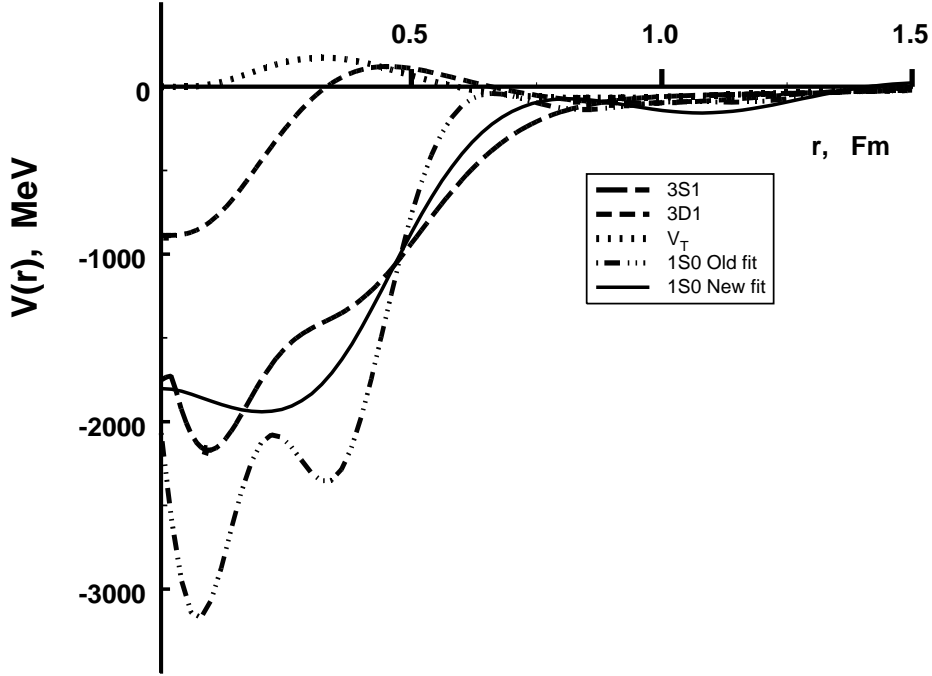}} \caption{Real parts of  NN potentials. Solid line $^1 S_0$ -one channel (inversion from new fit
of fig. 2). Dash-three points line $^1 S_0$ -one channel (inversion from old fit  of fig. 3). Two bound channels long
dashed line $V_{CS}(r)$ ($^3 S_1$), short dashed line $V_{CD}(r)$ ($^3 D_1$), dotted line $V_{tens}(r)$.}
\end{figure}

\begin{figure}[b] \epsfysize=100mm \centerline{
\epsfbox{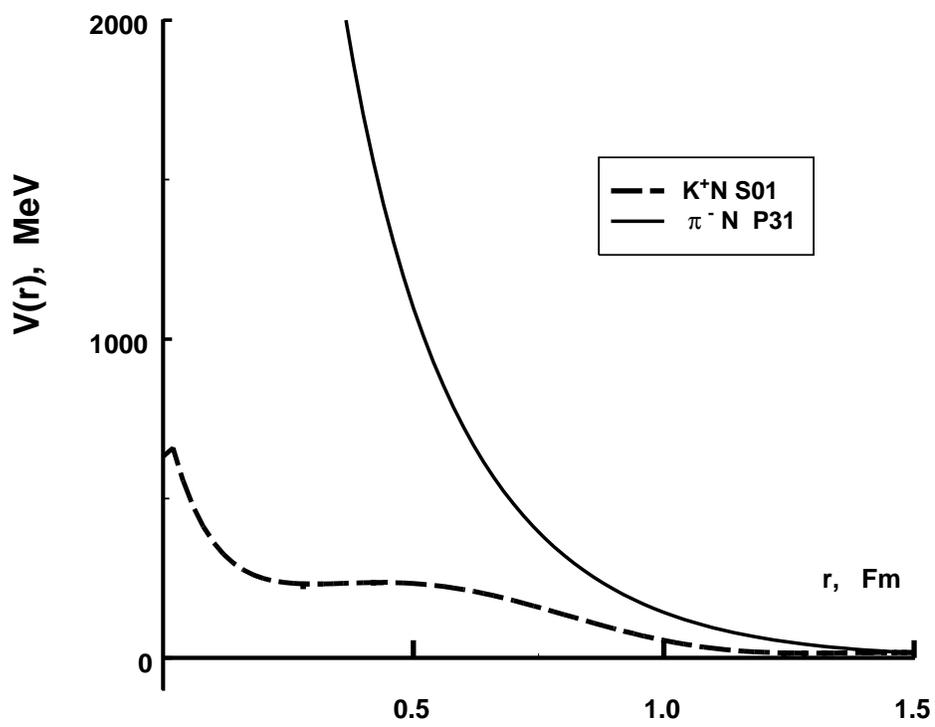}} \caption{Real parts of $\pi^{-}N$ $P31$ and $K^{+}N$ $S01$ potentials.}
\end{figure}

\begin{figure}[t]
\epsfysize=50mm \centerline{ \epsfbox{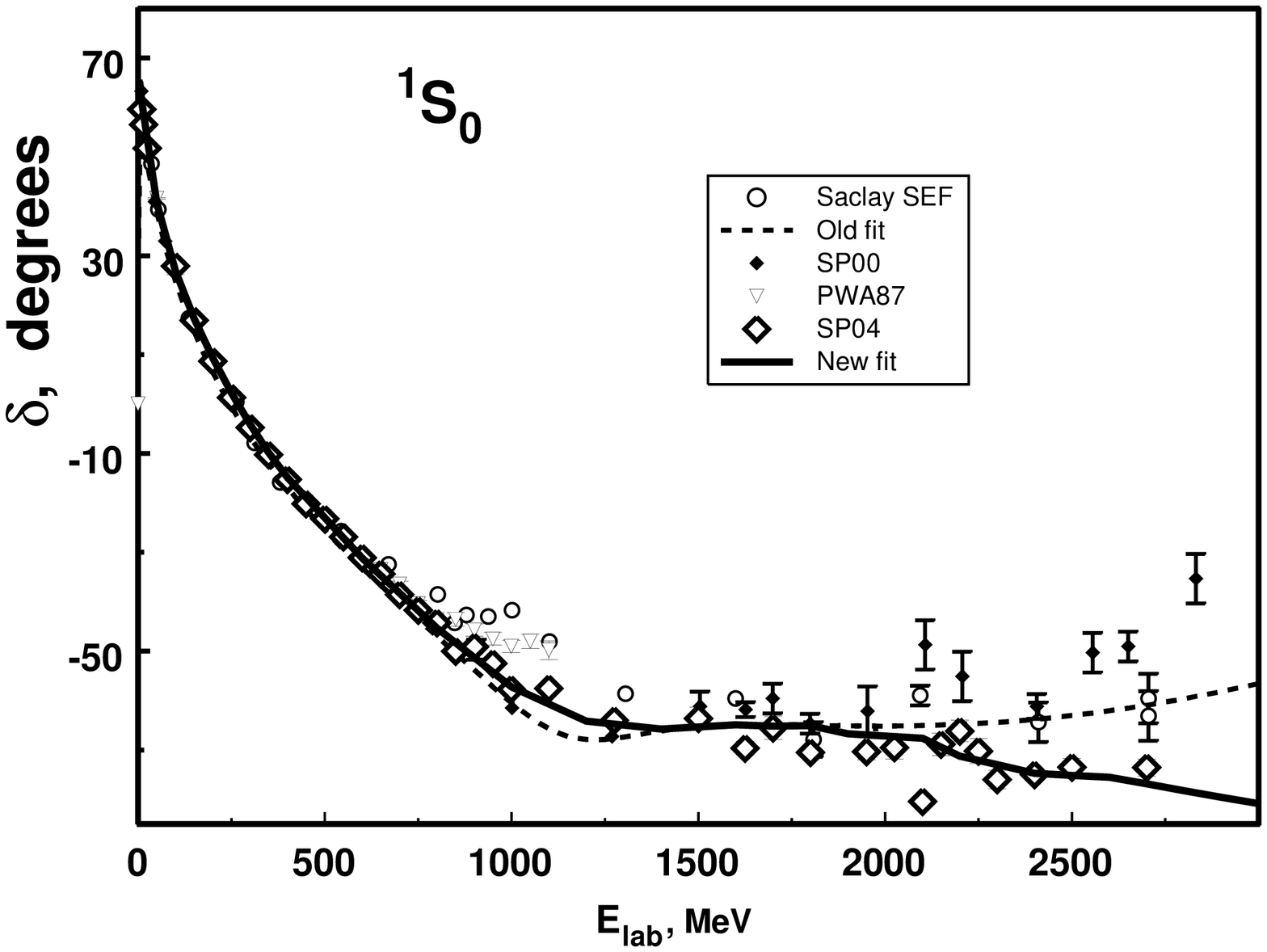} \epsfysize=50mm \epsfbox{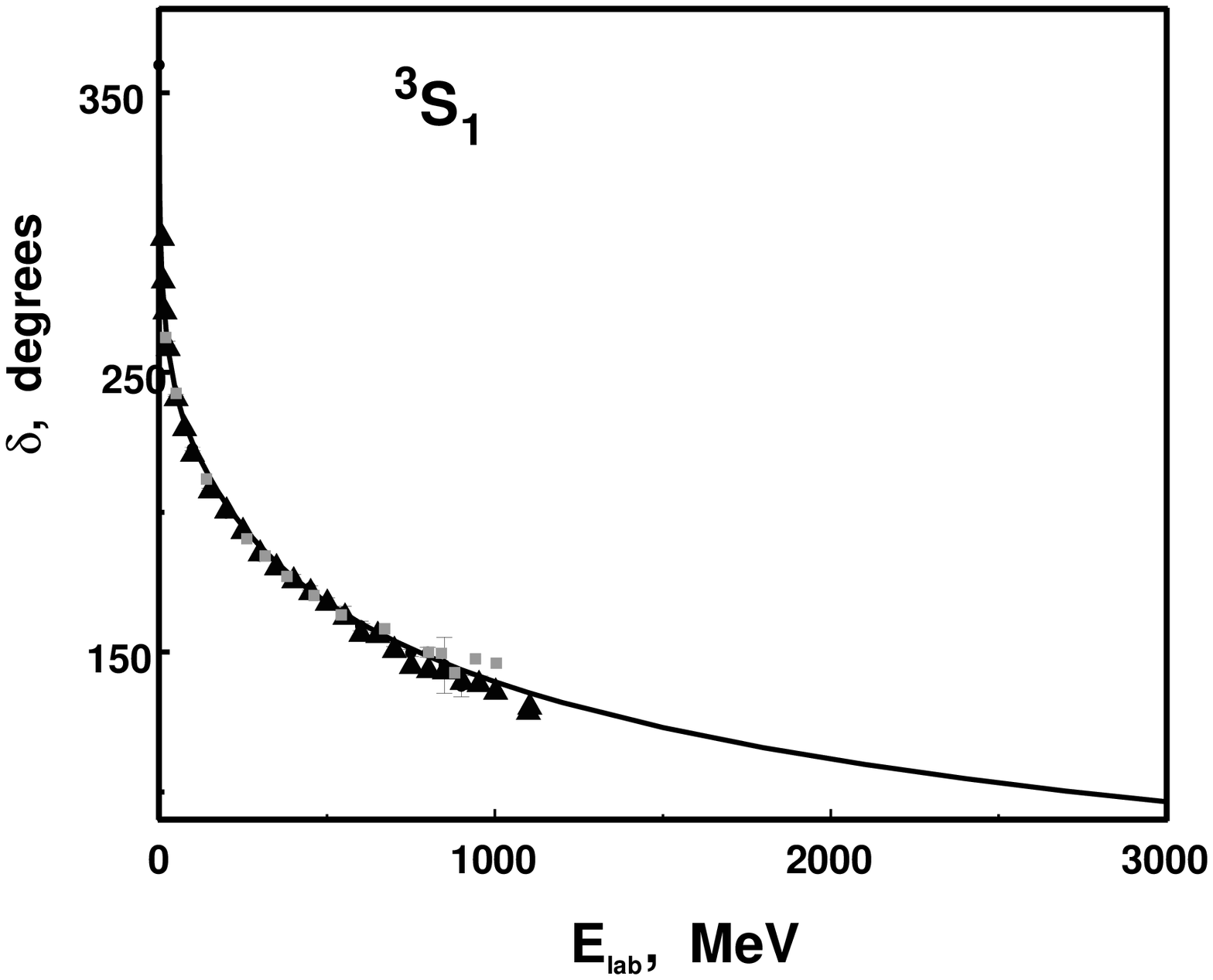}} \epsfysize=50mm \centerline{
\epsfbox{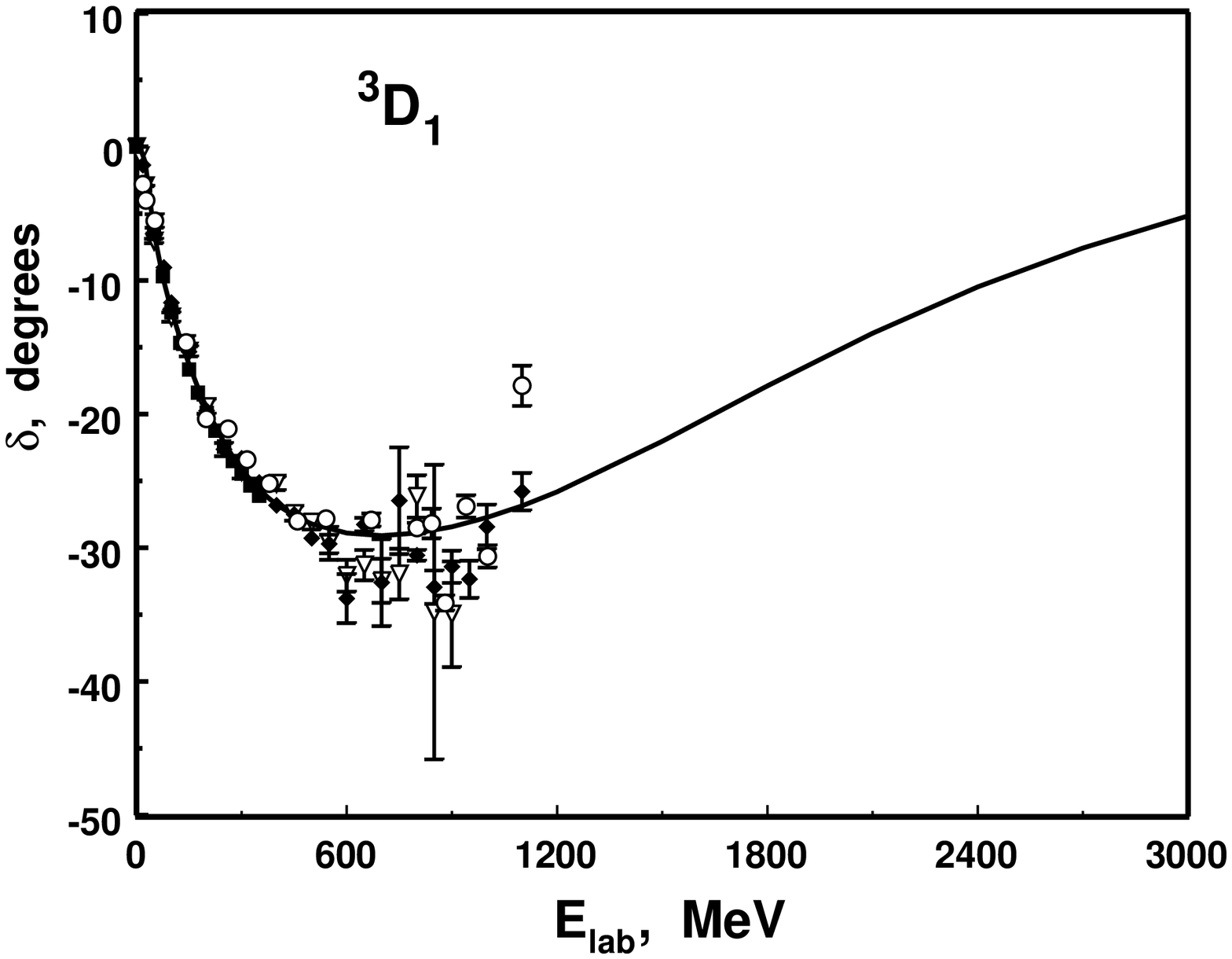} \epsfysize=50mm \epsfbox{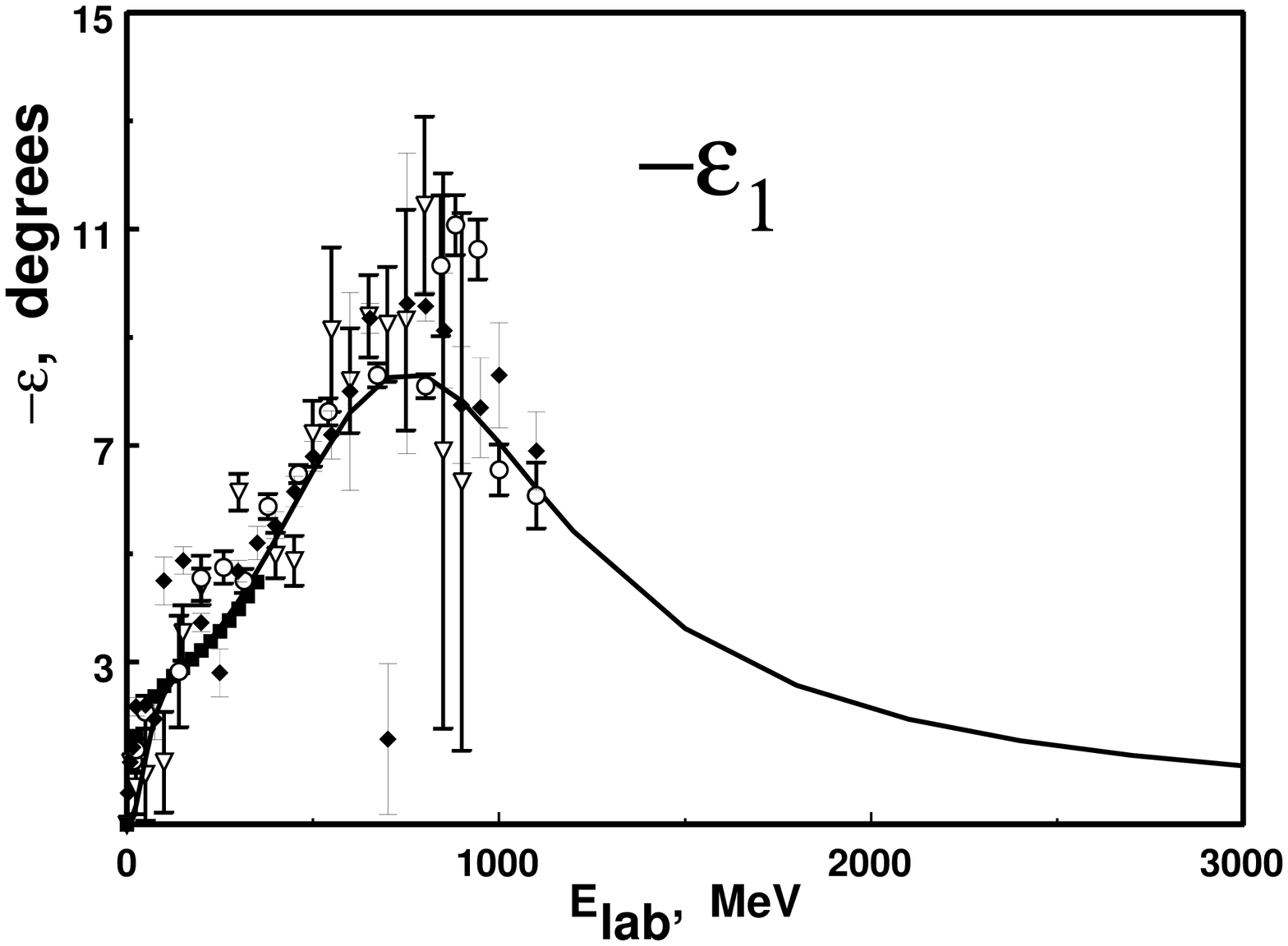}}\caption{Phase shifts and mixing parameter. Results for optical
potentials are indistinguishable from that as for real potentials. Solid lines, reconstructed Moscow potential; dashed
lines reconstructed repulsive core potential. The phase shift analysis  data are from \cite{DataScat,cite}. For $S$
waves the original data set from \cite{DataScat,cite} is raised 180 degrees up. To leave the $S$-matrix unchanged we
must then change the sign of the mixing parameter $\epsilon_1$ for the Moscow potential. SP00 and SP04 data are from
\cite{cite}.}
\end{figure}
\begin{figure}[t]
\epsfysize=50mm \centerline{ \epsfbox{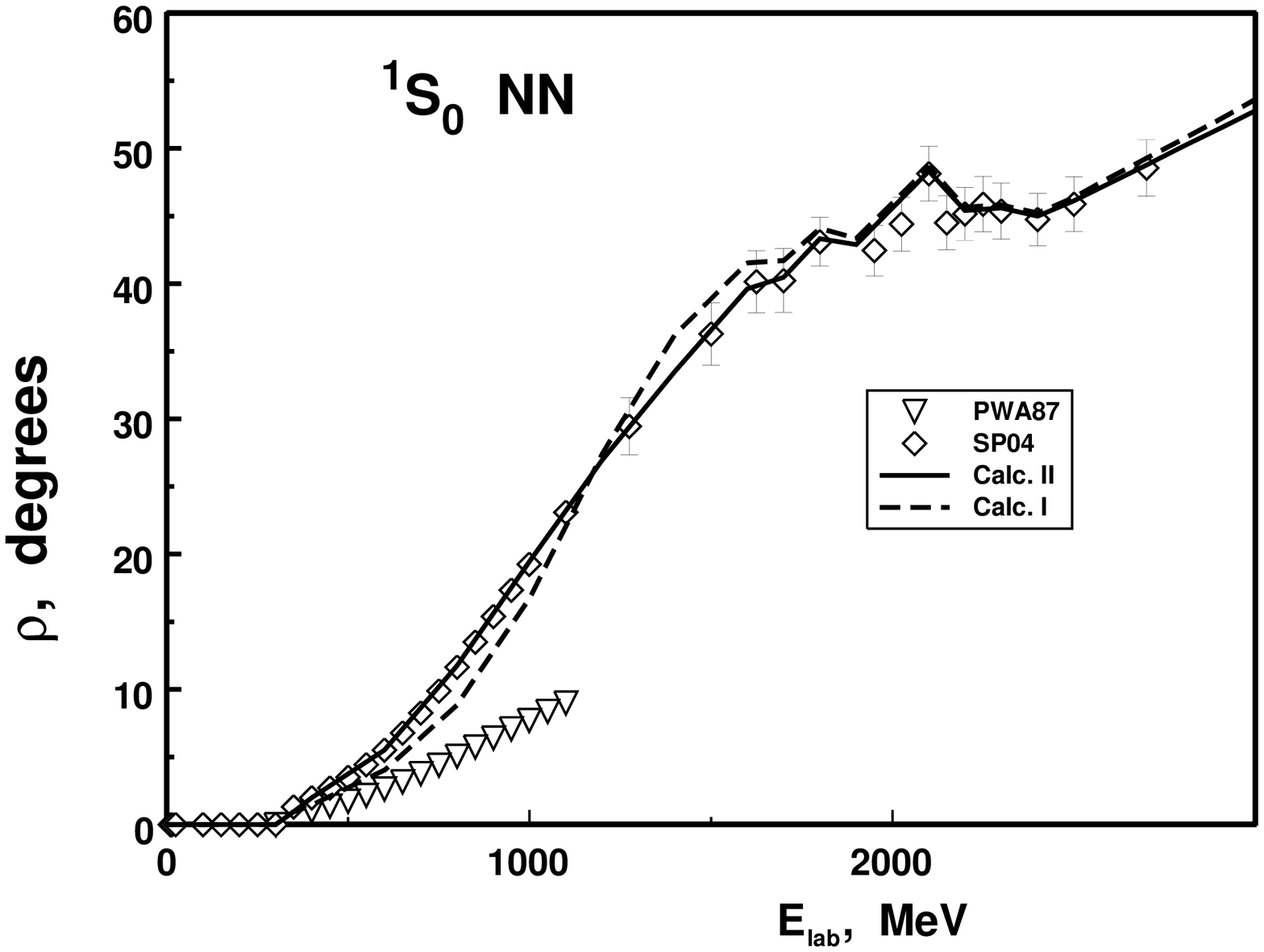} \epsfysize=50mm \epsfbox{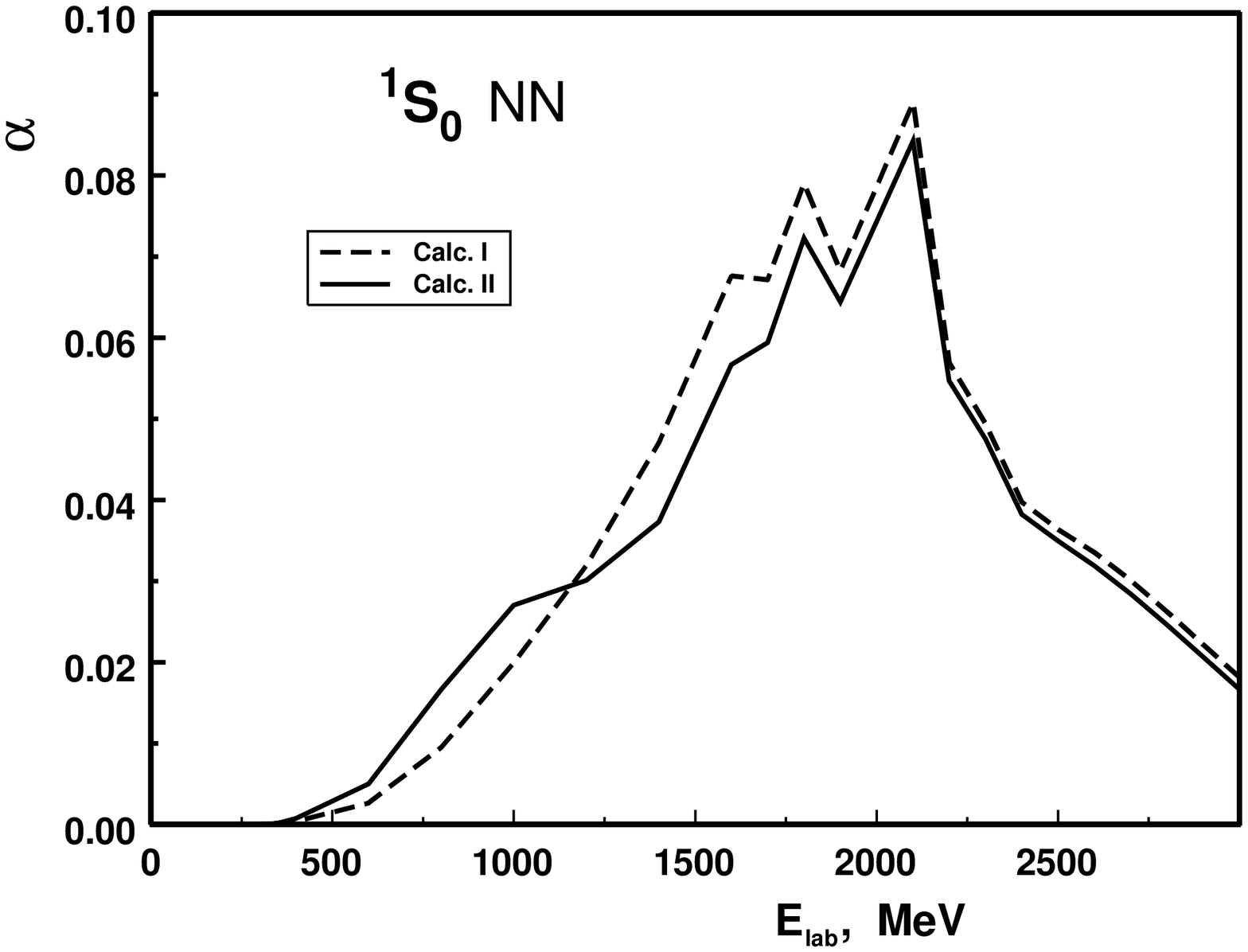} } \caption{Left: The
inelasticity parameters. The phase shift analysis data (circles) are from \cite{DataScat}.  Right: parameter $\alpha$.
 }
\end{figure}

\begin{figure}[t]
\epsfysize=100mm \centerline{\epsfbox{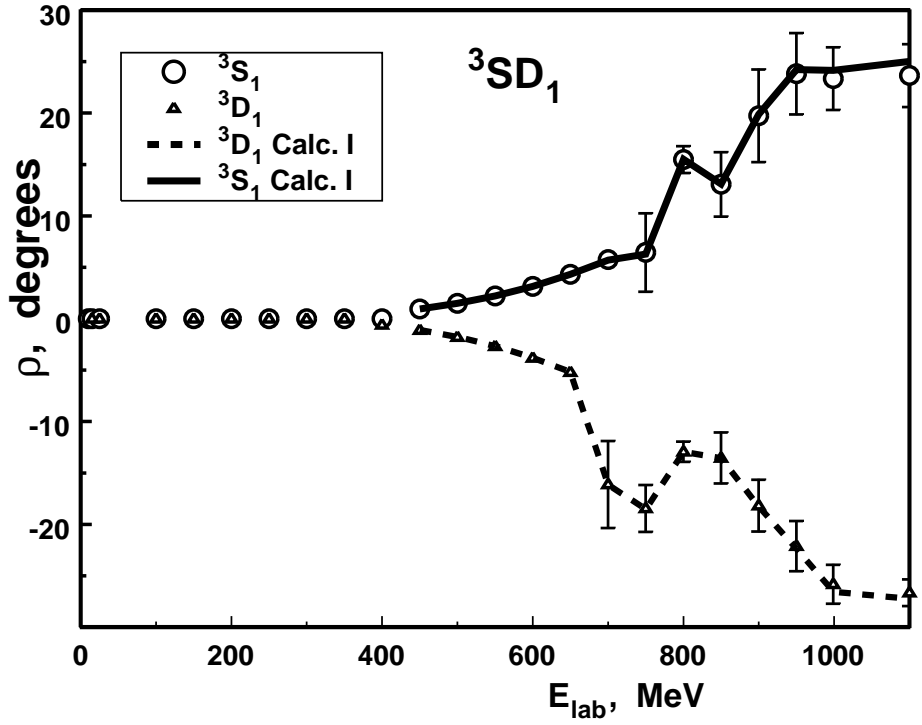}} \caption{The inelasticity parameters ${\rho}$ for $^3 SD_1$ waves.
The phase shift analysis data (circles) are from \cite{cite}.}
\end{figure}

\begin{figure}[t]
\epsfysize=100mm \centerline{\epsfbox{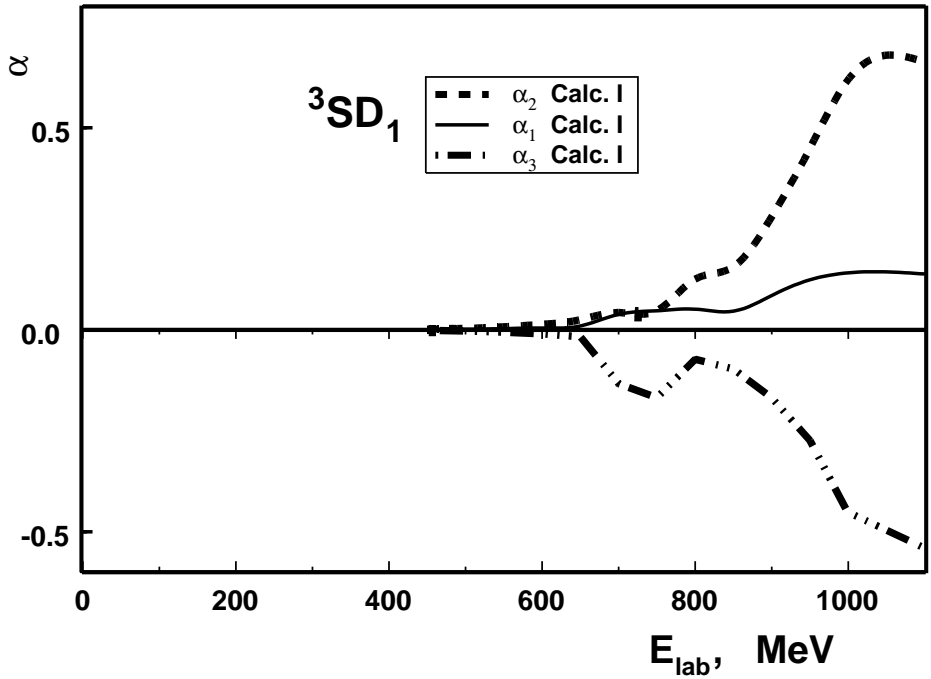}} \caption{ Parameters $\alpha_{i}$ for $^3 SD_1$ NN potential
predicted by eqs. (\ref{CouplDelta2}) and (\ref{Simpler}). Solid line, $\alpha_{1}$; dashed line with circles,
$\alpha_{2}$; dotted line, $\alpha_{3}$.}
\end{figure}

\begin{figure}[t]
\epsfysize=100mm \centerline{\epsfbox{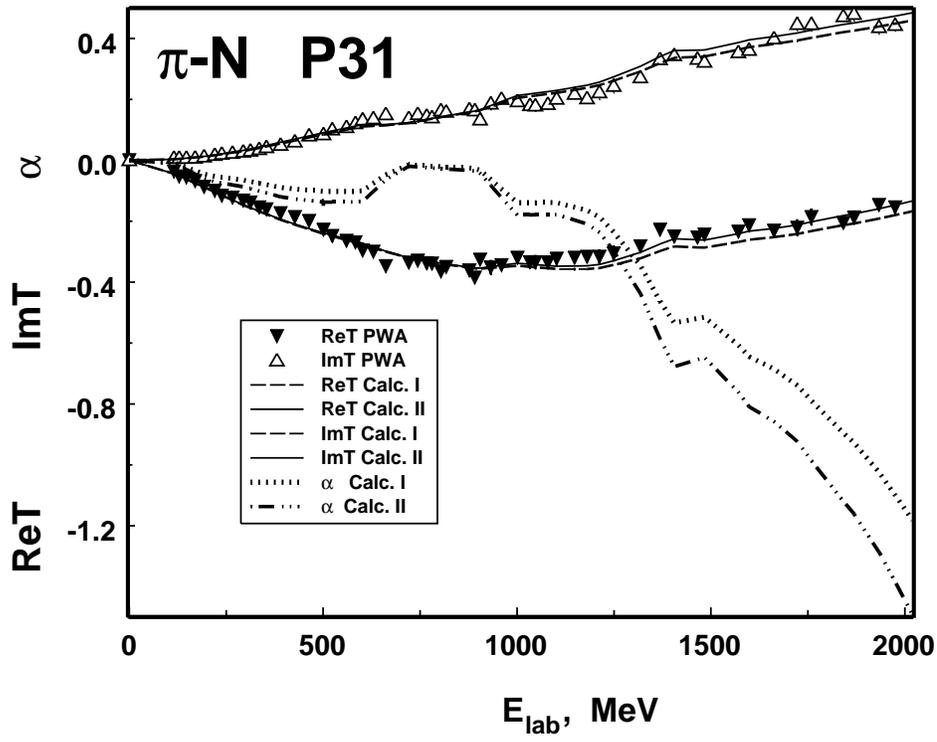}} \caption{$T$-matrix and $\alpha$ for $P31$ $\pi^{-} N$, data are
from \cite{cite}. }
\end{figure}
\begin{figure}[t]
\epsfysize=100mm \centerline{\epsfbox{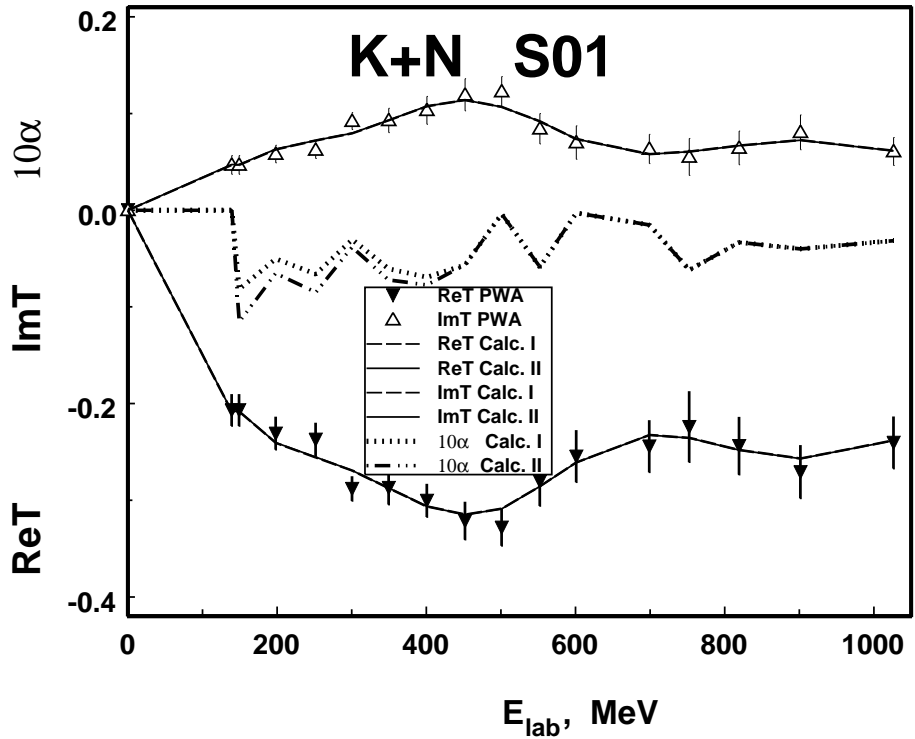}} \caption{$T$-matrix and $\alpha$  for  $S01$ $K^{+}N$, data are from
\cite{cite}. }
\end{figure}

\end{document}